# Genetic Studies of Physiological Traits with Their Application to Sleep Apnea

Dung-Yang Lee,[1] Craig Hanis,[2] Graeme Bell,[3] David Aguilar,[4] Brian Cade,[5] Jennifer Below,[2] Momiao Xiong,[1,2,*]

[1]Division of Biostatistics, University of Texas School of Public Health, Houston, TX 77030, USA

[2]Human Genetics Center, University of Texas School of Public Health, Houston, TX 77030, USA

[3]Department of Human Genetics and Department of Medicine, University of Chicago, IL 60637, USA

[4]Department of Cardiology, Baylor College of Medicine, Houston, TX 77030, USA

[5]Department of Medicine, Harvard University, Boston, MA 02115, USA

[*]Correspondence: Momiao.Xiong@uth.tmc.edu

**Abstract**

Advances of modern sensing and sequencing technologies generate a deluge of high dimensional space-temporal physiological and next-generation sequencing (NGS) data. Physiological traits are observed either as continuous random functions, or on a dense grid and referred to as function-valued traits. Both physiological and NGS data are highly correlated data with their inherent order, spacing, and functional nature which are ignored by traditional summary-based univariate and multivariate regression methods designed for quantitative genetic analysis of scalar trait and common variants. To capture morphological and dynamic features of the data and utilize their dependent structure, we propose a functional linear model (FLM) in which a trait curve is modeled as a response function, the genetic variation in a genomic region or gene is modeled as a functional predictor, and the genetic effects are modeled as a function of both time and genomic position (FLMF) for genetic analysis of function-valued trait with both GWAS and NGS data. By extensive simulations, we demonstrate that the FLMF has the correct type 1 error rates and much higher power to detect association than the existing methods. The FLMF is applied to sleep data from Starr County health studies where oxygen saturation were measured in 22,670 seconds on average for 833 individuals. We found 65 genes that were significantly associated with oxygen saturation functional trait with P-values ranging from 2.40E-06 to 2.53E-21. The results clearly demonstrate that the FLMF substantially outperforms the traditional genetic models with scalar trait.

## Introduction

Physiological traits such as electrocardiogram (ECG), phonocardiogram (PCG), seimocardiography (SCG), and oxygen saturation levels provide important information on the health status of humans and can be used to monitor, diagnose and manage diseases. For example, ECG is a measurement of the electrical activity of the heart muscle obtained from the surface of skin. It measures the rate and regularity of heart beats. ECG is the most commonly performed cardiac test and of great clinical value.[1] It provides valuable information on the biological processes and current state of the heart, and can be used for diagnosis of arrhythmias[2,3] , myocardial infarction and other cardiovascular diseases[4]. Oxygen saturation level which are proportional to the reduction in airflow cause total or partial reduction in respiration of the Sleep Apnea-Hypopnea Syndrome (SAHS) during sleep.[5] SAHS is a risk factor for cardiac and cerebral infarct, high arterial pressure, arrhythmias, and in general, several dysfunctions of the cardiorespiratory system. The physiological traits can be described as a mathematical function of time $t$ or other indexes which are often referred to as function-valued traits.[6]

Although many physiological traits are measured as a function, the widely used methods for genetic studies of physiological traits in humans are the same as that for the traditional single-valued quantitative traits where a single number is taken as a quantitative trait. These methods use summary statistic to measure or represent physiological traits. For example, heart rate (HR), the P-R interval, QRS complex duration, QT, and QTc interval are often used as a trait in genetic analysis of ECG.[7-9] Physiological traits are time dependent and dynamic in nature. They are repeatedly measured at multiple time points and often described by functions or curves. The temporal pattern of genetic control for physiological traits should be compared across different stages of development.[10] To capture the morphological shape and dynamic features of the physiological traits, methods that analyze all dynamic time points (traits) jointly and are often referred to as function valued QTL analysis have recently developed.[11,12] Quantitative genetic analysis of function-valued traits enhance our understanding the genetic control of the whole dynamic process of the traits and improve the statistical power to detect QTL. They explore the inherent order, spacing and functional nature of the data often ignored by traditional marginal genetic models.[13]

Although the genetic study of quantitative traits has been widely performed and experienced extensive technical development, the quantitative genetic analysis of function-valued trait is comparatively less development.[14] To our knowledge, no statistical methods have been developed for genetic analysis of function-valued traits with next-generation sequencing (NGS) data. In the past few years we have witnessed the rapid development of novel statistical methods for association studies using NGS data.[15-19] But, these methods might not be appropriate for genetic analysis of function-valued trait. The quantitative genetic analysis of rare variants for function-valued trait remains challenging. To meet the challenge, we propose the function linear model with functional responses and functional predictors (FLMF) for quantitative genetic analysis of function-valued traits with NGS data. In the FLMF, the time varying values of physiological trait are taken as a functional response and the genotype profile across a genomic region or a gene can be modeled as a function of genomic location. The FLMF has several remarkable features. First, the FLMF accounts for the continuous change in traits and preserves the intrinsic structure and all the positional-level genetic information. Second, the FLMF simultaneously utilize both correlation information among the trait at different times and among all variants in a genomic region. Third, the multicolinearity problems in the FLMF which may be presented in both trait and genetic variation is alleviated. Forth, the FLMF expand both trait function and genotype function in terms of orthogonal eigenfunction, which lead to substantial dimension reduction.

To evaluate its performance for genetic association analysis of function-valued trait, we use large scale simulations to calculate the type I error rates of the FLMF for testing the association of a genomic region (gene) with the function-valued trait. We evaluate the power of six statistical models. For the time course trait data, we considered the FLM with both trait function and genotype function, the multiple linear model for multiple phenotypes and simple regression for multiple phenotypes. For the cross-sectional data, we considered the FLM with scalar trait and genotype function, multiple linear model for single phenotype and simple regression for single phenotype. To further evaluate its performance, the FLMF is applied to oxygen saturation studies in Starr County, Texas where a total of 35,280 measurements were taken over a night and a total of 406,299 SNPs in 22,670 genes was typed for 833 individuals with Mexican Americans origin from Starr County. A program for implementing the developed FLMF for quantitative genetic analysis of function-valued trait can be downloaded from our website http://www.sph.uth.tmc.edu/hgc/faculty/xiong/index.htm

## Material and Methods

### Functional linear models with both functional response and predictors for genetic studies of temporal or space quantitative Trait

For the convenience of discussion, temporal or space trait is referred to as a functional trait. We consider a temporal trait $y_i(t), t \in \mathrm{T} = [0, T]$ of the $i$-th individual which varies in time and a genomic region (or gene) $[a, b]$. Let be a genomic position in the region $S = [a, b]$. Define a genotype profile $X_i(s)$ of the i-th individual as

$$X_i(s) = \begin{cases} 2P_m(\mathrm{s}), & \mathrm{MM} \\ P_m(\mathrm{s}) - \mathrm{P_M}(\mathrm{s}), & \mathrm{Mm} \\ -2P_M(\mathrm{s}), & \mathrm{mm} \end{cases}$$

where M and m are two alleles of the marker at the genomic position $s$, $P_M(s)$ and $P_m(s)$ are the frequencies of the alleles M and m, respectively. A functional linear model with both functional response and predictors is defined as

$$y_i(t) = W_i^T \alpha(t) + \int_S x_i(s)\beta(s,t)ds + \varepsilon_i(t), \; t = t_1,...,t_T, i = 1,...,n, \qquad (1)$$

where $W_i$ is a vector of covariates, $\alpha(t) = [\alpha_1(t),...,\alpha_d(t)]^T$ is a vector of effects associated with the covariates, $\beta(s,t)$ is a genetic additive effect function in the genomic position $s$ and time $t$, and $\varepsilon_i(t)$ is the residual function of the noise and unexplained effect for $i^{th}$ individual. Let $\eta(t) = [\eta_1(t),...,\eta_k(t)]^T$ is a vector of basis functions. To transform the functional linear model (1) into the standard multivariate linear mode, we consider the following Fourier functional expansions for the trait function $y_i(t)$, effect functions $\alpha(t), \beta(s,t)$ and genotype function $x_i(s)$:

$$y_i(t) = \sum_{l=1}^{k_y} y_{il}\eta_l(t) = y_i^T\eta(t), \; \alpha_j(t) = \sum_{l=1}^{k_y} \alpha_{jl}\eta_l(t) = \alpha_j^T\eta(t), \alpha(t) = \alpha\eta(t), \; \eta(t) = [\eta_1(t),...,\eta_{k_y}(t)]^T,$$

$\alpha$ is a $d \times k_y$ matrix, $\beta(s,t) = \sum_{k=1}^{k_\beta}\sum_{l=1}^{k_y} b_{kl}\theta_k(s)\eta_l(t) = \theta^T(s)B\eta(t)$, $\theta(s) = [\theta_1(s),...,\theta_{k_b}(s)]^T$,

$B = (b_{kl})_{k_\beta \times k_y}$ is a matrix of expansion coefficients of the genetic additive effect function,

$\varepsilon_i(t) = \sum_{l=1}^{k_y} \varepsilon_{il}\eta_l(t) = \varepsilon_i^T\eta(t)$, $\varepsilon_i = [\varepsilon_{i1},...,\varepsilon_{ik_y}]^T$. The integral $\int_S x_i(s)\beta(s,t)ds$ can be expanded as

$$\int_S x_i(s)\beta(s,t)ds = \int_S x_i(s)\theta^T(s)dsB\eta(t) = x_i^T B\eta(t),$$

where

$$x_i = [\int_T x_i(s)\theta_1(s)ds,...,\int_T x_i(s)\theta_{k_\beta}(s)ds]^T = [x_{i1},...,x_{ik_\beta}]^T .$$

Substituting these expansions into equation (1), we obtain

$$y_i^T \eta(T) = W_i^T \alpha\eta(t) + x_i^T B\eta(t) + \varepsilon_i^T \eta(t), i = 1,...,n . \quad (2)$$

Since equation (2) should hold for all $t$, we must have

$$y_i^T = W_i^T \alpha + x_i^T B + \varepsilon_i^T , i = 1,...,n. \quad (3)$$

The model (3) is a standard linear model. Instead of using the observed data as the values of the response and predictor variables, we use their expansion coefficients as the values of the response and predictor variables in the linear model (3). Equation (3) can be further written in a matrix form:

$$\begin{aligned} Y &= W\alpha + XB + \varepsilon, \\ &= A\gamma + \varepsilon, \end{aligned} \quad (4)$$

where $Y = [y_1,..., y_n]^T$ ,$W = [W_1,...,W_n]^T$ , $X = [x_1,..., x_n]^T$ , $\varepsilon = [\varepsilon_1,...,\varepsilon_n]^T$ , $A = [W, X]$ and $\gamma = [\alpha^T , B^T ]^T$ .

The least square estimates of the parameter vector $\gamma$ is given by

$$\hat{\gamma} = (A^T A)^{-1} A^T Y. \quad (5)$$

Using equation (5) we obtain the genetic additive effect function:

$$\beta(s,t) = \theta^T (s) B\eta(t) . \quad (6)$$

**Test Statistics**

An essential problem in genetic studies of the functional quantitative trait is to test the association of a genomic region with the functional quantitative trait. Formally, we investigate the problem of testing the following hypothesis:

$$H_0 : \beta(s,t) = 0 \ , \qquad \forall s \in S, t \in \mathrm{T} \tag{7}$$

against

$$H_a : \beta(s,t) \neq 0 \ .$$

If the genetic effect function $\beta(s,t)$ is expanded in terms of the basic functions:

$$\beta(s,t) = \theta^T(s) B \eta(t) ,$$

then, testing the null hypothesis $H_0$ in equation (7) is equivalent to testing the hypothesis:

$$H_0 : B = 0 \ . \tag{8}$$

To derive the test statistic, we first calculate variance of the estimated expansion coefficient matrix of the genetic additive effect function. Let vec denote the vector operation. Then, from equation (5), we have

$$\begin{aligned} vec(\hat{\gamma}) &= [I_{k_y} \otimes (A^T A)^{-1} A^T] vec(y) \\ &= vec(\gamma) + [I_{k_y} \otimes (A^T A)^{-1} A^T] vec(\varepsilon). \end{aligned} \tag{9}$$

Note that

$$\mathrm{var}(vec(\varepsilon)) = \Sigma_{k_y} \otimes I_n , \tag{10}$$

where

$$\Sigma_{k_y} = \begin{bmatrix} \sigma_{11} & \cdots & \sigma_{1k_y} \\ \vdots & \vdots & \vdots \\ \sigma_{k_y 1} & \cdots & \sigma_{k_y k_y} \end{bmatrix}. \qquad (11)$$

From equation (3), we obtain

$$y_{ij} = \sum_{k=1}^{d} W_{ik}\alpha_{kj} + \sum_{k=1}^{k_y} x_{ik}b_{kj} + \varepsilon_{ij}, i = 1,...,n, j = 1,...,k_y \ . \qquad (12)$$

Variance $\sigma_{ul}$ can be estimated by

$$\sigma_{ul} = \frac{1}{nk_y - dk_y - k_y k_\beta} \sum_{i=1}^{n} (y_{iu} - \sum_{k=1}^{d} W_{ik}\hat{\alpha}_{ku} - \sum_{k=1}^{k_y} x_{ik}b_{ku})(y_{il} - \sum_{k=1}^{d} W_{ik}\hat{\alpha}_{kl} - \sum_{k=1}^{k_y} x_{ik}b_{kl}), u = 1,...,k_y, l = 1,..,k_y$$

.

Then, $\mathrm{var}(vec(\hat{\gamma}))$ is given by

$$\begin{aligned} \mathrm{var}(vec(\hat{\gamma})) &= [I_n \otimes (A^T A)^{-1} A^T]\mathrm{cov}(vec(\varepsilon))[I_n \otimes A(A^T A)^{-1}] \\ &= [I_n \otimes (A^T A)^{-1} A^T][\Sigma_{k_y} \otimes I_n][I_n \otimes A(A^T A)^{-1}] \\ &= \Sigma_{k_y} \otimes (A^T A)^{-1}. \end{aligned} \qquad (13)$$

Let

$\hat{b} = vec(\hat{B})$ and $\Lambda$ be the matrix that are obtained from the last $k_y k_\beta$ row and $k_y k_\beta$ columns of the matrix $\mathrm{var}(vec(\hat{\gamma}))$. We can define the following statistic for testing the association of a genomic region with the functional trait:

$$T_F = \hat{b}^T \Lambda^{-1} \hat{b}. \qquad (14)$$

Under the null hypothesis of no association, the statistic $T_F$ will be distributed as a central $\chi^2_{(k_y k_\beta)}$ distribution.

## Results

### Null Distribution of Test Statistics

In the previous section, we have shown that the test statistics $T_F$ are asymptotically distributed as a central $\chi^2_{(k_y k_\beta)}$ distribution. To examine the validity of this statement, we performed a series of simulation studies to compare their empirical levels with the nominal ones.

We calculated the type I error rates for rare alleles, and both rare and common alleles. We assumed the following model to generate a functional quantitative trait for type 1 error calculations:

$$y_i(t_j) = \mu + \varepsilon_i(t_j),$$

where $y_i(t_j)$ is the trait value of $i^{th}$ individual at the time $t_j$, $\mu$ is a constant for all $i$ and $t_j$, $\varepsilon_i(t_j)$ is the error term of $i^{th}$ individual at the time $t_j$, and this error term is generated by independent standard Brownian motion.

We first considered both common and rare variants, i.e., entire allelic spectrum of variants. We generated 1,000,000 chromosomes by resampling from 2,225 individuals with variants in five genes (*CDC2L1*, *GBP3*, *IQGAP3*, *TNN*, *ACTN2*) selected from the NHLBI's Exome Sequencing Project (ESP) . The five genes included 461 SNPs. The number of sampled individuals from populations of 1,000,000 chromosomes ranged from 1,000 to 2,000. The time points taking trait measurement for type 1 error calculations were 15, 20, 30 and 40. A total of 5,000 simulations were repeated.  Tables 1 and 2 summarized the average type I error rates of the test statistics for testing the association of rare variants (MAF < 0.05) and all common and rare variants over five genes, respectively, at the nominal levels α=0.05, α=0.01 and α=0.001. Tables 1 and 2 showed that in general, the type I error rates of the test statistics in the functional quantitative trait analysis were not appreciably different from the nominal alpha levels.

**Power Evaluation**

To evaluate the performance of the functional linear models with both functional response and predictors for testing the association of a genomic region with a functional quantitative trait, we used simulated data to estimate their power to detect a true association. A true functional quantitative genetic model is given as follows. Consider $L$ trait loci that are located at the genomic positions $s_1,...,s_L$. Let $A_s$ be a risk allele at the $s^{th}$ trait locus. Let $t_j$ be the $j$ -th time point when the trait measurement is taken. The following multiple linear regression is used as an additive genetic model for a quantitative trait:

$$y_i(t_j) = \mu + \sum_{s=1}^{L} x_{is} b_s(t_j) + \varepsilon_i(t_j),$$

where $y_i(t_j)$ is the trait value of $i^{th}$ individual measured in the time $t_j$, $\mu$ is an overall mean, $x_{is}$ is an indicator variable for the genotype of $i^{th}$ individual at the $s^{th}$ trait locus, $b_s(t_j)$ is the genetic additive effect of the SNP at the $s^{th}$ trait locus and the time $t_j$, the error term $\varepsilon_i(t_j)$ is generated by independent standard Brownian motion process. The genetic effect $b_s(t_j)$ is modeled as $b_s(t_j) = b_s b(t_j)$, where $b(t_j) = 10^{-6} e^t$. We considered four genetic models for $b_s$: additive, dominant, recessive and multiplicative. The relative risks across all variant sites are assumed to be equal and the variants were assumed to influence the trait independently (i.e. no epistasis). Let $f_0 = 1$ be a baseline penetrance that is defined as the contribution of the wild genotype to the trait variation and $r$ be a risk parameter. The genetic additive effects for the four trait models are defined as follows:

dominant model: $b_s = (1 - P_s)(r-1) f_0$, recessive model: $b_s = P_s(r-1) f_0$,

additive model: $b_s = (P_s + r - 1) f_0$ and multiplicative model: $b_s = (rP_s + 1 - P_s)(r-1) f_0$, where $P_s$ is the frequency of the risk allele located at the genomic position $s$.

For power comparisons, we also consider cross-section trait models. The genetic effects for the cross-section trait models is defined as average of the genetic effect function over the time where the phenotype values were measured at 20 time points: $\bar{b}_s = b_s b(t_{med})$, where $b(t_{med})$ is the median of the function of $b(t_j), j = 1,...,20$. The trait value for the cross-sectional model is generated by

$$y_i = \mu + \sum_{s=1}^{L} x_{is}\bar{b}_s + \varepsilon_i .$$

We generate 100,000 individuals by resampling from 2,225 individuals of European origin with variants in gene *TNN* (88 rare variants and 18 common variants) selected from ESP dataset. We randomly selected 10% of the variants as risk variants. A total of 1,000 individuals for the dominant, additive and multiplicative trait models and 2,000 individuals for the recessive trait model were sampled from the populations. A total of 1,000 simulations were repeated for the power calculation. We compared the power of six methods. For the time course trait data, we considered the FLM with both trait function and genotype function, the multivariate regression for multiple phenotypes and simple regression for multiple phenotypes. For the cross-sectional data, we considered the FLM with scalar trait and genotype function, multivariate regression for single phenotype and simple regression for single phenotype.

We compare the power curves of FLM with cross-sectional model, Multivariate Regression Model and Simple Linear Regression Model in this study. We repeat 1,000 simulations for all the comparisons. Also, we assuming that all variances are independently and equally influence the trait. That is we assume there are no interactions happen in those variances.

Figures 1, 2 and Supplementary Figures 1 and 2 plot the power curves of six statistic models: the functional linear model with both functional response and predictors for function-valued trait (FLMF), the multiple linear model for function-valued trait (MLMF), the simple regression model for function-valued trait (SRGF), the functional linear model with scalar response and functional response for cross section marginal genetic model (FLMC), multiple linear model for cross section marginal genetic model and simple regression for cross section marginal genetic model (SRGC) for testing association of rare variants in the genomic region under dominant, additive, multiplicative, and recessive models, respectively. These power curves are a function of the risk parameter at the significance level $\alpha = 0.05$. Several features emerged from these figures. First, the power of the FLMF was the highest. Except for the recessive models, the FLMF could still detect association of a gene with the function-valued trait even using sample sizes of 1,000. Second, power difference between the FLMF and other five models was substantial. Third, the power of simple regression for both function-valued trait and cross section marginal model (SRGF and SRGC) was extremely low. In most scenario, the simple regression had not power to detect association. Forth, in general, the power of tests using function-valued approach was higher than that using traditional cross section approach.

Now we study the power of six models for testing the association of both common and rare variants. Figures 3 and 4, and Supplementary Figures 3 and 4 plotted the power of six models for testing association of 18 common variants and 88 rare variants in the genomic region as a function of the risk parameter at the significance level $\alpha = 0.05$ under the additive, multiplicative, dominant and recessive models, respectively. Again, we observed the same pattern of their power for testing association of all common and rare variants as that for testing association of rare variants. The power of all statistics under additive model was higher than that under other three trait models. Also, we observed that when risk parameter exceeds 1.6 the power of the FLMC and MLMC was higher than the power of MLMF. In any scenarios, the FLMF had the highest power to detect association among six models. This demonstrated that the FLMF can be our best choice in quantitative trait association studies no matter whether the variants are common or rare. As we expected, the power of the tests for both common and rare variants was higher than that for testing association of rare variant only.

**Application to Real Data Examples**

To further evaluate its performance, the FLMF was applied to oxygen saturation studies in Starr County, Texas. The oxygen saturation signals were measured by seconds. A total of 35,280 measurements were taken over a night. Oxygen saturation provides important information on the sleep quality of the obstructive sleep apnea.[20] A total of 406,299 SNPs in 22,670 genes was typed for 833 individuals with Mexican Americans origin from Starr County. Since the FLMF requires to expand genotype function in terms of eigenfunction, which need to have at least 3 SNPs in the gene, we excludes the gene with only one or two SNPs in it. The left total number of genes for analysis was 17,258. Therefore, the P-value for declaring significance after applying the Bonferroni correction for multiple tests was $2.90 \times 10^{-6}$. Distributions of gender, age, sleeping time and BMI were summarized in Table S1. To reduce the number of measurements included in the analysis, we used the mean of the oxygen saturation in every 10 seconds as the trait values.

SNPs in 5*kb* flanking region of the gene are assumed to be belong to the gene. To ensure the numerical stability, we used single value decomposition to calculate the inverse of the matrix.[21] We selected the number of single values such that it can account for 99% of total variation.

To examine the behavior of the FLMF, we plotted QQ plot of the test (Figure 5) where P-values were calculated after adjusting for sex, age and BMI in the model. The QQ plots showed that the false positive rate of the FLMF for detection of association with oxygen saturation trait is controlled. In total, we identified 65 genes that were significantly associated with oxygen saturation function-valued trait with P-values ranging from $2.4 \times 10^{-6}$ to $2.5 \times 10^{-21}$ (Table 3). Instead of using whole oxygen saturation curve as functional response, we also used their mean as scalar response variable and applied the FLMC for testing association. In Table 3, we included the P-values of using the FLMC for testing association of a gene with the mean oxygen saturation overnight. To compare with other methods for association analysis of function-valued traits, we provided Table S2 in which we also listed minimum of P-values of 65 significant genes over all observed time period which were calculated using MLM and SRG for each time point.

Several remarkable features were observed from this real data analysis. First, the FLMF utilizes the merits of taking both phenotype and genotype as functions. It decomposes time varying phenotype function into orthogonal eigenfunctions of time and position varying genotype function into orthogonal eigenfunctions of genomic position. The FLMF reduces the dimensions due to both phenotype variation and genotype variation (only a few eigenfunctions are used to model variation), which in turn increases statistical power of the test. Table 3 clearly demonstrated that P-values calculated by the FLMF were much smaller than that by the FLMC. The models for genotype variation within a gene for the FLMF and FLMC are the same. Only difference between the FLMF and FLMC is how to model the phenotype. The FLMF model the phenotypes as curve or function while the FLMC models the phenotypes as its mean or a scalar. This real data example showed that the function-value (time course data) approach can achieve much stronger significance than the scalar value (cross section study) approach. Second, to further illustrate that the function-valued statistical methods can be more powerful than the traditional quantitative genetic analysis, we presented Table S2 showing that the P-values of the FLMF were smaller than the minimum of P-values of the MLM and SRG over all observed time interval at night. Third, genetic variants in a gene might make only mild contribution to the oxygen saturation variation at individual time point, these genetic variants may show significant association with the oxygen saturation curve as shown in Figure S5 where the P-value for testing the association of gene *ANKLE1* with the oxygen saturation curve using the FLMF was $2.51\times10^{-14}$ and the P-values for the tests using the MLM at the individual time point ranges from $6.52\times10^{-7}$ to 0.9265. There were a total of 3,528 time points. We observed a total of 188 time points with P-value < 0.05 when using the MLM to test association at the individual time points. None of the 3,528 tests showed strong evidence of association, but indeed we observed strong association of the gene *ANKLE1* with the oxygen saturation curve due to using all information about correlation and continuity of underlying structure of phenotype function. Fourth, unlike traditional quantitative genetic analysis where a single constant P-value for the test is calculated, in the genetic analysis of function-valued trait we can observe the time varying P-values. To illustrate this, we plotted Figure S6 showing the P-values of the MLMT for testing the association of all SNPs within the gene *TMEM50B* with the oxygen saturation at each time point over night as a function of time $t$ .There was the rapid changes of P-value of the MLM test over time. We observed two peaks showing significant association with the oxygen saturation.

At most times during the night the genetic variation in the gene *TMEM50B* did not have big impact on the variation of the oxygen saturation.

The genetic effect in the FLMF is characterized by its spatiotemporal pattern. The genetic effect is a function of both time $t$ and genomic position $s$. Similar to the concept of probability density function in the probability theory, the genetic effect function is viewed as the average genetic effect in a unit interval of time (or index value) and the genomic region. The genetic effect function is more interpretable than the scattered spatiotemporal genetic effect points of the SNPs within the gene. It often consists of several peaks and valleys where the values at the peak of the genetic effect function are the synthesized genetic effects of the individual SNPs in the region due to the correlation between the peak and nearby time and SNPs. To illustrate this, we plotted Figure 6 showing the genetic effect function $\beta(s,t)$ of the gene *TMEM50B* (P-value < $9.59\times10^{-9}$) as a function of time and the genomic position in the FLMF model. The genetic effect function surface $\beta(s,t)$ provide full and detailed spatiotemporal information on how and what genetic variants affect the development of biological process, which will lead to new biological insight.

Among 65 significant genes, five genes *RELB, MAFF, EEF1A1, CDC42EP5* and *COMMD7* are involved in pro-inflammatory NF-kB signaling pathway.[23-27] Pro-inflammatory NF-kB signaling pathway plays an important molecular role in linking between sleep apnea, obesity, and inflammation. Targeting the NF-κB pathway will ameliorate the metabolic dysregulation involved in obstructive sleep apnea (OSA).[28] OSA is associated with increased risk of development of obesity, hypertension, diabetes and heart diseases. It was reported that genes *GIPC1, LYRM1, LHX2* and *RELB* were associated with obesity,[29-32] genes *EEF1A1* and *GFP1* were involved in diabetes,[33-35] and genes *GIPC1, TMEM57, GGT1, LDLRAP1, COMMD7,TPM4* played an important in heart disease.[36-40]

## Discussion

The current paradigm for genetic studies of physiological trait is summary-based quantitative genetic analysis where function-valued phenotypes are represented as summary statistics. However, physiological traits are repeated measurements of dynamic biological processes and best to be represented as a mathematic function. Summary statistics cannot capture morphological and dynamic features of physiological traits. Summary-based statistical methods for quantitative genetic analysis of physiological traits are lack of power to detect association of genetic variants with the whole biological process. To overcome the limitations of summary-based statistical approaches, the function-valued methods for quantitative genetic analysis have rapidly developed recently. Although they use the time order and spacing of the data and take continuous change in traits of interests into account, these function-valued methods for quantitative genetic analysis which are mainly designed for common variants, test association variant by variant and do not use order and spacing information of the genomic data. The current function-valued methods for genetic analysis which only model the traits as a mathematical function, but still model the genetic variant variation as univariate or multivariate variables are not well suited for quantitative genetic analysis of physiological traits with NGS data. To address the critical barriers in genetic analysis of function-valued trait, we proposed function linear models with both phenotype function and genotype function for quantitative genetic analysis of physiological traits with NGS data where we not only model the trait as a function, but also model the genetic variation across a genomic region as a function. We take a genomic region or a gene as a basic unit of analysis and collectively test the association of a genomic region or gene with a function-valued trait. By large simulations and real data analysis we demonstrate the merits and limitations of the proposed new paradigm of association analysis for function-valued trait.

The new approach uses both all trait information in the measured time interval and all genetic information in the genome region to collectively test association of all genetic variants within the regions with function-valued trait. In the FLMF model, we first separately expand the trait function in an interval and the genotype function in a genomic region (gene) in terms of orthonormal basis functions. The trait morphological and dynamic information across the time interval and genetic information across all variants in the genomic region including all single variant variation and their linkage disequilibrium are compressed into sets of expansion coefficients, one for the trait values and one for the genotype values. These expansion coefficients in the functional data analysis are referred to as functional principal component scores (FPC scores). The multivariate multiple regression is used to model the relationships between the FPC scores for the trait function and FPC scores for the genotype function. We use the compressed genetic information to globally test association of the genomic region or the gene with the function-valued trait. The FLMF is a natural extension of multivariate multiple regression. By large simulations and real data analysis, we showed that the proposed FLMF substantially increased the power and dramatically reduced the data noises.

The most widely used statistical methods for genetic studies of function-valued trait are originally designed for testing association of common variants with the function-valued traits. They are lack of power to test association of rare variants. The developed FLMF can efficiently test the association of the entire allelic spectrum of variants with function-valued trait.

Unlike simple and multiple regressions discarding a large amount of information due to using limited numbers to summarize the data, the FLM F preserves the intrinsic correlation structure in the trait and all the positional-level genetic information. The space-ordering of the trait dynamics and genetic variation data is a central feature in the FLMF. Both the neighboring trait values and genetic variants are linked. The trait value at one time point depends on the trait values at nearby time points. Similarly, the genotypes at one SNP are dependent on the genotypes at nearby SNPs. The most popular genetic analysis methods of function-valued trait will not account for the space-ordering of the data. The FLMF simultaneously employs individual trait values and their correlation, and genetic information of the individual variants and correlation information (LD) among all variants. It uses intrinsic functional dependence structure of the data and all available trait values in the time interval of interest and genetic information of the variants in the genomic region.

The genetic effect of function-valued trait is a function of time and genomic position of the SNP. It changes over time and genomic position. We observed its multiple peaks and valleys to indicate that there are specific SNPs showing significant association at particular times. Traditional summary-based univariate or multivariate regression analysis will smooth peaks and valley and hence lose power to detect association. Our real data analysis clearly demonstrates that a set of SNPs will jointly show strong association with function-valued trait, but they individually and at individuate time make only mild contribution to the association.

The trait values in a time interval and genetic variant data in a genomic region which often have strong correlation generate multicolinearity and high dimensionality which other methods are often unable to deal with efficiently. In the FLMF, the trait function and the genetic variant functions  are expended in terms of orthogonal or closely to orthogonal functions. The component coefficients in the expansion will, in general, not be linearly dependent. Therefore, the multicolinearity problem in the FLMF is alleviated.

Different numbers of measurements of trait values at different time points and genotype at different genomic position exist for each individual. The FLMF use these data to fit the curves that are used to test association, despite the sampling differences among individuals. Therefore, the FLMF can efficiently deal with missing data which often happen for the function-valued traits and NGS data.

Although the genetic analysis of function-valued traits has remarkable feature it suffers severe limitations due to the cost of additional measurements and unfamiliarity with function data analysis. The advances of current wireless and communication technologies facilitate a new generation of unobtrusive, portable, and ubiquitous health monitoring systems such as wearable ECG and other physiological sensors for continuous patient assessment and more personalized health care.[41] Cheap wearable and wireless sensors and a Personalized Wearable Monitoring System will generate rich physiological datasets (function-valued datasets). We can expect that more and more genetic studies of physiological (function-valued) traits will be performed in the near future.

Next-generation sequencing technologies will identify ten millions of genetic variants across the human genome and modern sensors will generate hundreds of thousands or even millions of values of physiological trait. Such extremely high-dimensional data that are full of noise and missing data pose fascinating statistical and computational challenges in genetic analysis of physiological traits. Transition of analysis from low dimensional data to extremely high dimensional data demands changes in statistical methods from multivariate data analysis to functional data analysis. In the past decade we have witnessed the emergence of the functional data analysis as an exciting research area of statistics which provides powerful and informative tools for the analysis of various types of high dimensional data including both physiological trait variation and genomic variation. Our theoretic results, real oxygen saturation genetic data analysis and simulations showed that the FLMF for genetic analysis of function-valued trait is able to fully explore all of the information contained in the phenotype and genotype data, efficiently utilize the merits of both point-by-point and joint analyses while overcome their limitations. Therefore, the FLMF is one of the choices in quantitative genetic analysis with NGS data. Emergence of sensing, wireless communication and sequencing technology, and application of the genomic continuum model and functional data analysis is expected to open a new era for quantitative genetic studies. The results in this paper are preliminary. The purpose of this paper is to stimulate further discussions regarding great challenges we are facing in the quantitative genetic studies of high dimensional phenotypic and genomic data produced by modern sensors and next-generation sequencing.

## Supplemental Data

Supplemental data included four Figures and two Tables.

## Acknowledgments

The project described was supported by Grant 1R01AR057120–01 and 1R01HL106034-01, from the National Institutes of Health and NHLBI, respectively.

**Web Resources**

A program for implementing the developed FLMF for quantitative genetic analysis of function-valued trait can be downloaded from our website http://www.sph.uth.tmc.edu/hgc/faculty/xiong/index.htm

**Figure Titles and Legends**

**Figure 1.** The power curves as a function of risk parameter of six models: the functional linear model with both functional response and predictors for function-valued trait (FLMF), the multiple linear model for function-valued trait (MLMF), the simple regression model for function-valued trait (SRGF), the functional linear model with scalar response and functional response for cross section marginal genetic model (FLMC), multiple linear model for cross section marginal genetic model and simple regression for cross section marginal genetic model (SRGC) for testing association of rare variants in the genomic region under dominant model at the significance level, assuming a baseline penetrance of 1 and sample sizes of 1,000.

**Figure 2.** The power curves as a function of risk parameter of six models: the functional linear model with both functional response and predictors for function-valued trait (FLMF), the multiple linear model for function-valued trait (MLMF), the simple regression model for function-valued trait (SRGF), the functional linear model with scalar response and functional response for cross section marginal genetic model (FLMC), multiple linear model for cross section marginal genetic model and simple regression for cross section marginal genetic model (SRGC) for testing association of rare variants in the genomic region under additive model at the significance level, assuming a baseline penetrance of 1 and sample sizes of 1,000.

**Figure 3.** The power curves as a function of risk parameter of six models: the functional linear model with both functional response and predictors for function-valued trait (FLMF), the multiple linear model for function-valued trait (MLMF), the simple regression model for function-valued trait (SRGF), the functional linear model with scalar response and functional response for cross section marginal genetic model (FLMC), multiple linear model for cross section marginal genetic model and simple regression for cross section marginal genetic model

(SRGC) for testing association of both common and rare variants in the genomic region under dominant model at the significance level, assuming a baseline penetrance of 1 and sample sizes of 1,000.

**Figure 4.** The power curves as a function of risk parameter of six models: the functional linear model with both functional response and predictors for function-valued trait (FLMF), the multiple linear model for function-valued trait (MLMF), the simple regression model for function-valued trait (SRGF), the functional linear model with scalar response and functional response for cross section marginal genetic model (FLMC), multiple linear model for cross section marginal genetic model and simple regression for cross section marginal genetic model (SRGC) for testing association of both common and rare variants in the genomic region under additive model at the significance level, assuming a baseline penetrance of 1 and sample sizes of 1,000.

**Figure 5.** QQ plot for P-values to test the association of 17,258 genes with the oxygen saturation function trait by the FLMF where *x* axis represents the expected –log10 (P-value) and *y* axis represents the observed –log10 (P-value).

**Figure 6.** The genetic effect of the gene *TMEM50B* as a function of time and the SNP located genomic position in the FLMF model.

## Supplementary Figures

**Figure S1.** The power curves as a function of risk parameter of six models: the functional linear model with both functional response and predictors for function-valued trait (FLMF), the multiple linear model for function-valued trait (MLMF), the simple regression model for function-valued trait (SRGF), the functional linear model with scalar response and functional response for cross section marginal genetic model (FLMC), multiple linear model for cross section marginal genetic model and simple regression for cross section marginal genetic model (SRGC) for testing association of rare variants in the genomic region under multiplicative model at the significance level, assuming a baseline penetrance of 1 and sample sizes of 1,000.

**Figure S2.** The power curves as a function of risk parameter of six models: the functional linear model with both functional response and predictors for function-valued trait (FLMF), the multiple linear model for function-valued trait (MLMF), the simple regression model for function-valued trait (SRGF), the functional linear model with scalar response and functional response for cross section marginal genetic model (FLMC), multiple linear model for cross section marginal genetic model and simple regression for cross section marginal genetic model (SRGC) for testing association of rare variants in the genomic region under recessive model at the significance level, assuming a baseline penetrance of 1 and sample sizes of 2,000.

**Figure S3.** The power curves as a function of risk parameter of six models: the functional linear model with both functional response and predictors for function-valued trait (FLMF), the multiple linear model for function-valued trait (MLMF), the simple regression model for function-valued trait (SRGF), the functional linear model with scalar response and functional response for cross section marginal genetic model (FLMC), multiple linear model for cross section marginal genetic model and simple regression for cross section marginal genetic model

(SRGC) for testing association of both common and rare variants in the genomic region under multiplicative model at the significance level, assuming a baseline penetrance of 1 and sample sizes of 1,000.

**Figure S4.** The power curves as a function of risk parameter of six models: the functional linear model with both functional response and predictors for function-valued trait (FLMF), the multiple linear model for function-valued trait (MLMF), the simple regression model for function-valued trait (SRGF), the functional linear model with scalar response and functional response for cross section marginal genetic model (FLMC), multiple linear model for cross section marginal genetic model and simple regression for cross section marginal genetic model (SRGC) for testing association of both and rare variants in the genomic region under recessive model at the significance level, assuming a baseline penetrance of 1 and sample sizes of 2,000.

**Figure S5.** –log10 P-value of the MLM for testing the association of gene *ANKLE1* with the oxygen saturation at each time point overnight as a function of time $t$ where the redline indicates the P-value declaring significance.

**Figure S6.** –log10 P-value of the MLM for testing the association of gene *TMEM50B* with the oxygen saturation at each time point overnight as a function of time $t$ where the redline indicates the P-value declaring significance.

Table 1. Average type 1 error rates of the statistics for testing association of a gene that consists of rare variants, MAF < 0.05, with a function quantitative trait over 5 genes.

| Time | Sample Size | 0.001 | 0.01 | 0.05 |
|---|---|---|---|---|
| 15 | 1000 | 0.00156 | 0.01232 | 0.05640 |
| | 1250 | 0.00088 | 0.01032 | 0.05320 |
| | 1500 | 0.00120 | 0.01264 | 0.05428 |
| | 1750 | 0.00152 | 0.01296 | 0.05484 |
| | 2000 | 0.00148 | 0.01144 | 0.05492 |
| 20 | 1000 | 0.00108 | 0.01164 | 0.05856 |
| | 1250 | 0.00088 | 0.01272 | 0.05876 |
| | 1500 | 0.00088 | 0.01152 | 0.05152 |
| | 1750 | 0.00144 | 0.01056 | 0.05508 |
| | 2000 | 0.00088 | 0.01068 | 0.05220 |
| 30 | 1000 | 0.00136 | 0.01200 | 0.05560 |
| | 1250 | 0.00112 | 0.01108 | 0.05232 |
| | 1500 | 0.00092 | 0.01032 | 0.05108 |
| | 1750 | 0.00136 | 0.01032 | 0.05204 |
| | 2000 | 0.00084 | 0.01020 | 0.05116 |
| 40 | 1000 | 0.00116 | 0.01184 | 0.05612 |
| | 1250 | 0.00128 | 0.01048 | 0.05344 |
| | 1500 | 0.00124 | 0.01144 | 0.05228 |
| | 1750 | 0.00096 | 0.01032 | 0.04876 |
| | 2000 | 0.00112 | 0.01100 | 0.05092 |

Time: the number of time points when taking trait measurement.

Table 2. Average type 1 error rates of the statistics for testing association of a gene that consists of all variants with a function quantitative trait over 5 genes.

| Time | Sample Size | 0.001 | 0.01 | 0.05 |
|---|---|---|---|---|
| 15 | 1000 | 0.0014 | 0.0130 | 0.0594 |
| | 1500 | 0.0010 | 0.0102 | 0.0552 |
| | 2000 | 0.0006 | 0.0116 | 0.0506 |
| 20 | 1000 | 0.0008 | 0.0100 | 0.0532 |
| | 1500 | 0.0014 | 0.0078 | 0.0490 |
| | 2000 | 0.0008 | 0.0090 | 0.0418 |
| 30 | 1250 | 0.0012 | 0.0138 | 0.0544 |
| | 1500 | 0.0006 | 0.0092 | 0.0458 |
| | 1750 | 0.0010 | 0.0076 | 0.0426 |
| 40 | 1250 | 0.0010 | 0.0126 | 0.0518 |
| | 1500 | 0.0008 | 0.0094 | 0.0482 |
| | 1750 | 0.0008 | 0.0086 | 0.0418 |

Time: the number of time points when taking trait measurement.

Table 3. P-values of 65 genes that were significantly associated with oxygen saturation trait.

| Gene | P-value | | Gene | P-value | | Gene | P-value | |
|---|---|---|---|---|---|---|---|---|
| | FLMF | FLMC | | FLMF | FLMC | | FLMF | FLMC |
| MAN1B1 | 2.53E-21 | 4.33E-01 | GIPC1 | 1.93E-09 | 7.01E-01 | CDC42EP5 | 7.88E-08 | 9.21E-01 |
| TMEM57 | 8.90E-18 | 8.50E-02 | CDC14C | 2.09E-09 | 1.97E-01 | MIR29C | 1.30E-07 | 9.65E-01 |
| OR5H15 | 1.28E-17 | 4.37E-01 | MIR4520B | 2.49E-09 | 2.64E-01 | LHX2 | 1.95E-07 | 7.85E-01 |
| PABPC4L | 2.66E-15 | 4.16E-01 | MIR4520A | 2.49E-09 | 9.33E-01 | ZNF284 | 2.09E-07 | 9.89E-01 |
| ANKLE1 | 2.51E-14 | 3.31E-01 | PROX2 | 3.63E-09 | 2.59E-01 | RBAKDN | 2.34E-07 | 9.63E-01 |
| TTI2 | 4.64E-14 | 9.27E-01 | MAFF | 4.04E-09 | 8.90E-01 | BAIAP2L2 | 2.44E-07 | 2.21E-01 |
| KRTAP4-7 | 1.67E-13 | 2.15E-01 | UQCRQ | 5.95E-09 | 4.54E-03 | P2RX5 | 2.92E-07 | 9.55E-01 |
| WDR90 | 1.73E-13 | 7.37E-02 | LYRM1 | 6.58E-09 | 1.58E-01 | P2RX5-TAX | 2.92E-07 | 9.90E-01 |
| ZER1 | 1.67E-12 | 3.15E-03 | ZFPM1 | 7.81E-09 | 9.58E-01 | RELB | 3.01E-07 | 7.20E-01 |
| DPH2 | 3.05E-12 | 8.86E-01 | TMEM50B | 9.59E-09 | 8.55E-01 | TREML3P | 5.42E-07 | 2.38E-01 |
| B9D2 | 3.43E-12 | 3.97E-01 | KCNK15 | 1.78E-08 | 5.45E-01 | TSPAN10 | 5.86E-07 | 7.50E-01 |
| GGT1 | 4.29E-12 | 5.68E-01 | SNORA41 | 1.99E-08 | 2.64E-01 | RPS16 | 6.17E-07 | 9.75E-01 |
| SGSH | 4.90E-12 | 5.86E-01 | EEF1B2 | 1.99E-08 | 6.19E-01 | GNLY | 8.09E-07 | 3.99E-01 |
| FAM211B | 1.03E-11 | 3.13E-01 | SNORD51 | 1.99E-08 | 1.84E-01 | LRRC48 | 8.17E-07 | 9.28E-01 |
| FBXO27 | 1.37E-11 | 3.41E-01 | LDLRAP1 | 2.01E-08 | 1.50E-02 | WSB1 | 9.64E-07 | 6.54E-01 |
| COA6 | 1.44E-11 | 8.04E-01 | NEK4 | 2.17E-08 | 5.00E-01 | GFPT1 | 1.09E-06 | 4.80E-01 |
| MAK16 | 2.66E-11 | 4.10E-01 | COMMD7 | 2.29E-08 | 2.56E-01 | MIR3677 | 1.13E-06 | 9.08E-02 |
| CDKN2AIP | 1.69E-10 | 3.25E-01 | EEF1A1 | 3.39E-08 | 5.44E-01 | MIR940 | 1.13E-06 | 8.26E-01 |
| RRM2 | 3.07E-10 | 4.79E-01 | MIR1-1 | 4.51E-08 | 8.76E-01 | TOE1 | 1.13E-06 | 2.97E-01 |
| DTX3L | 1.24E-09 | 8.80E-01 | HMGN4 | 4.63E-08 | 9.24E-01 | TMEM41A | 2.12E-06 | 4.95E-01 |
| C17orf75 | 1.41E-09 | 2.61E-01 | EVPLL | 5.48E-08 | 8.84E-01 | TPM4 | 2.40E-06 | 2.71E-01 |
| TAS2R5 | 1.72E-09 | 8.55E-01 | C22orf26 | 5.60E-08 | 6.66E-01 | | | |

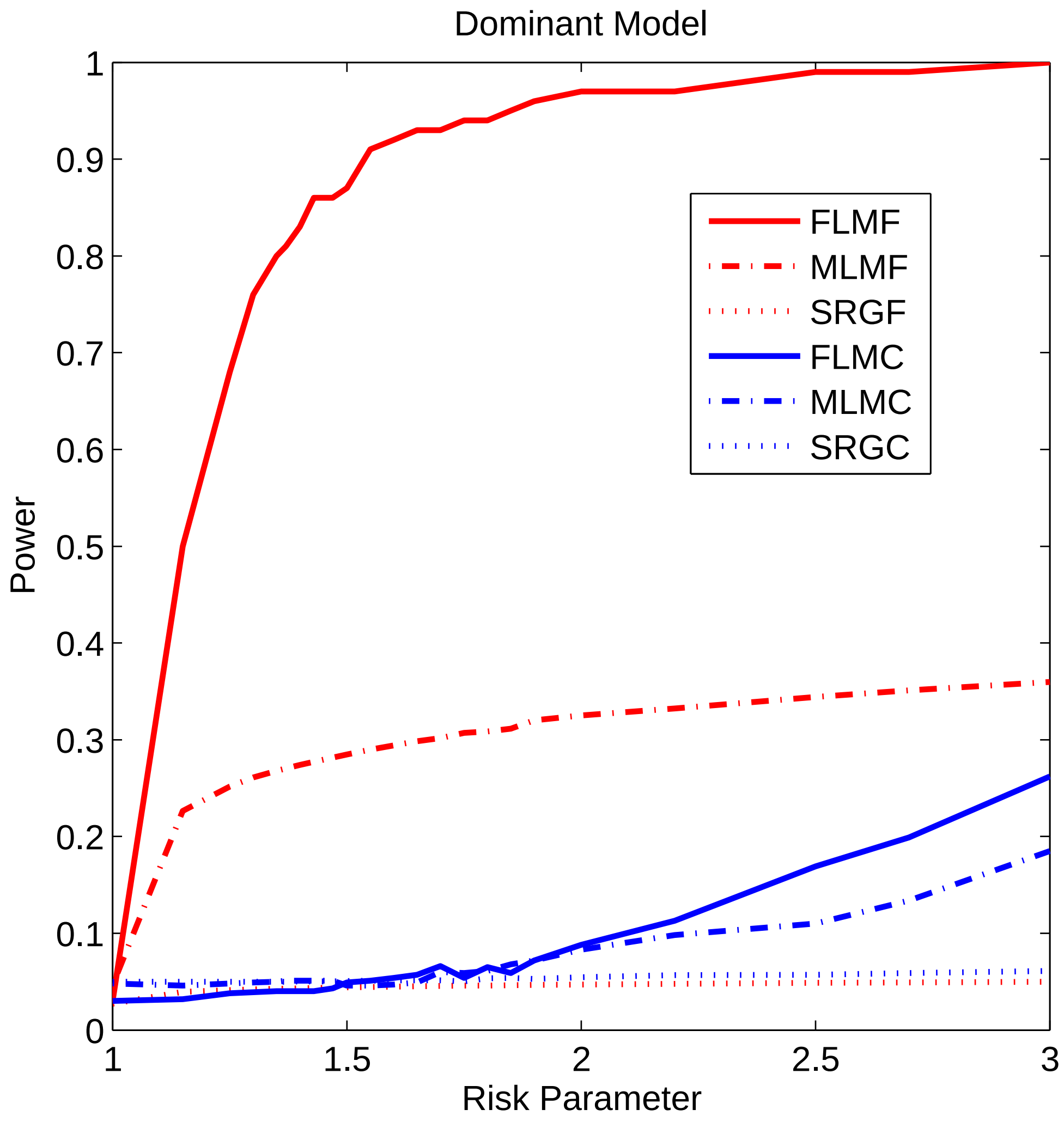

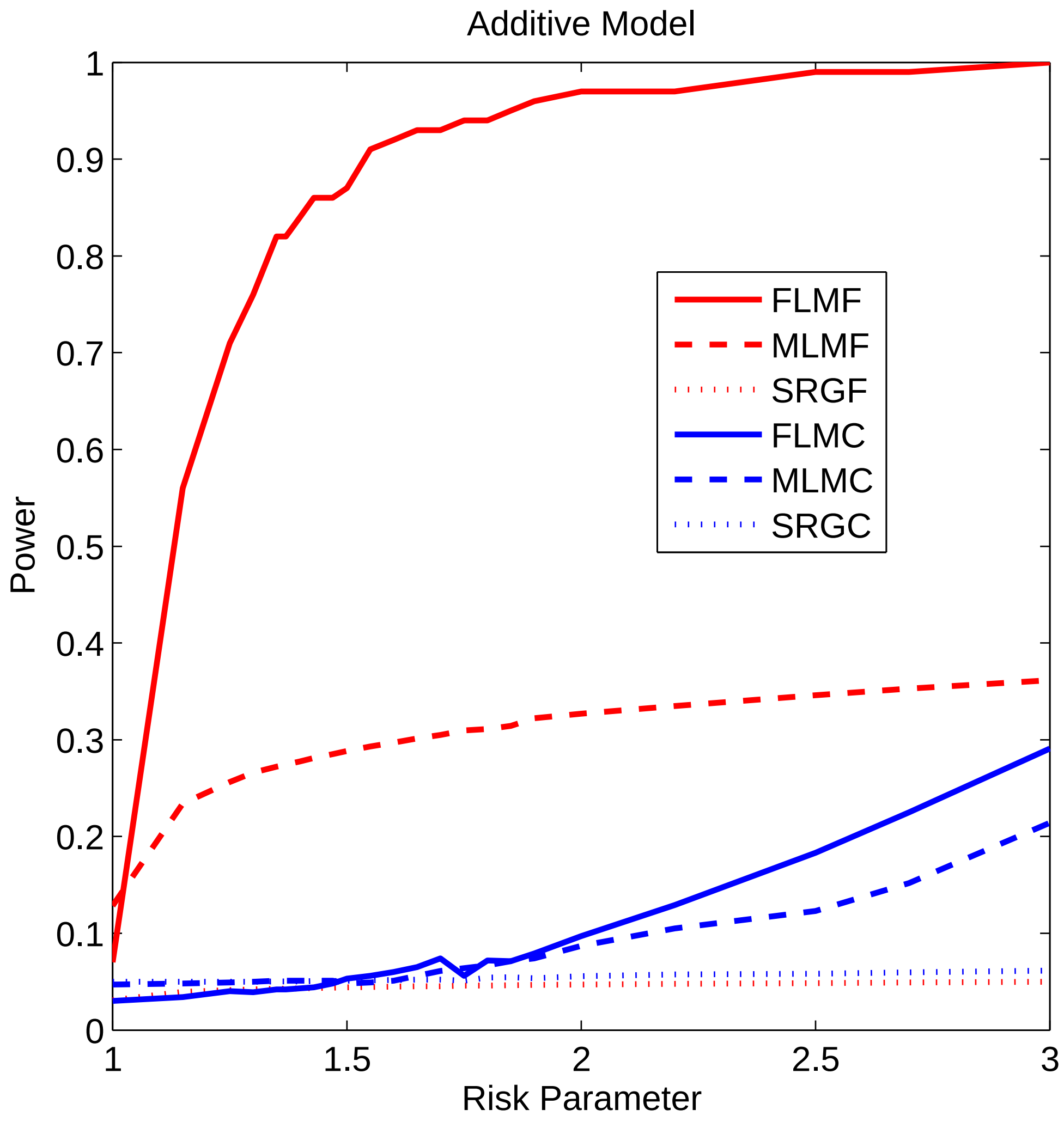

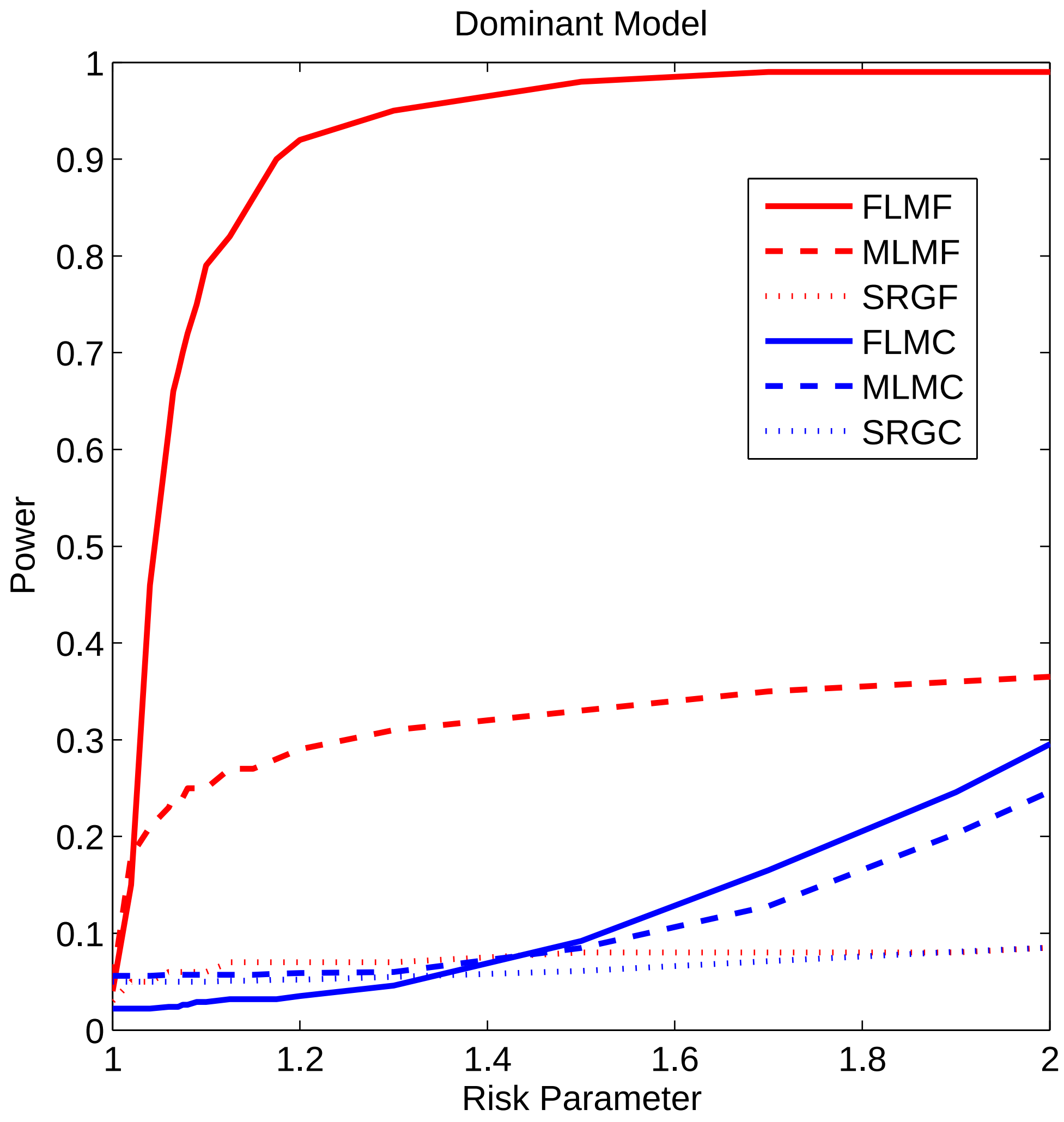

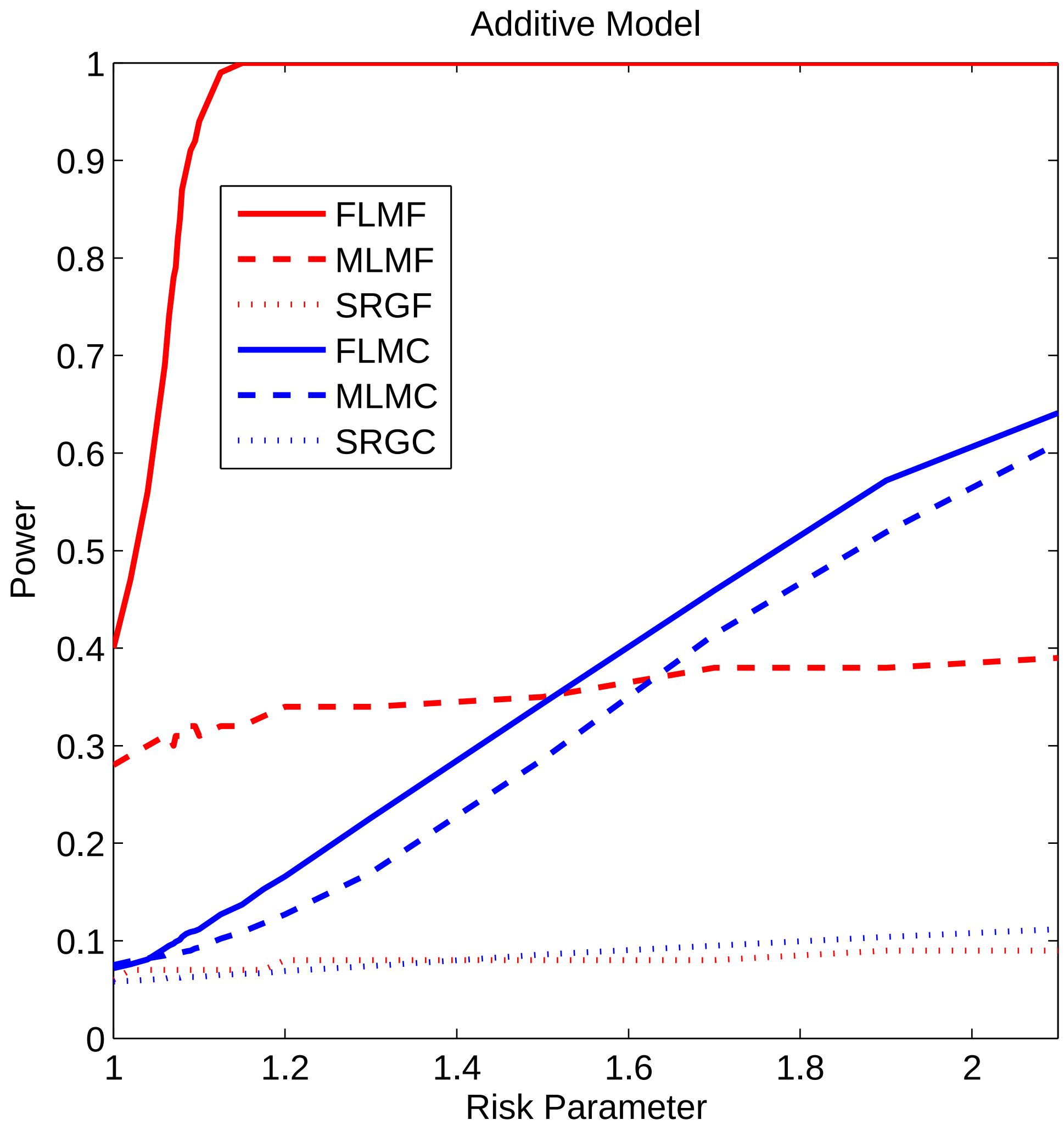

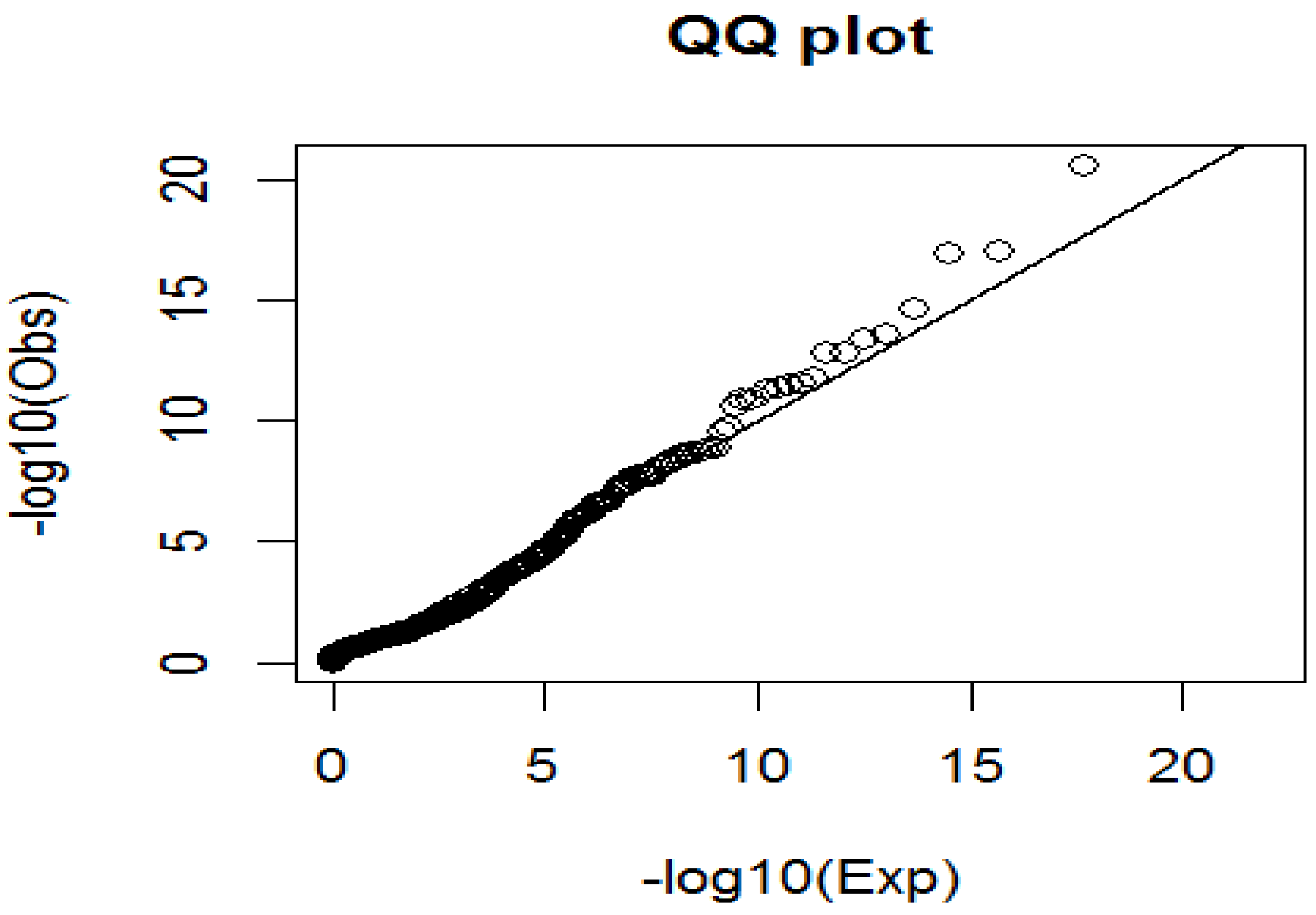

**B(s,t)**

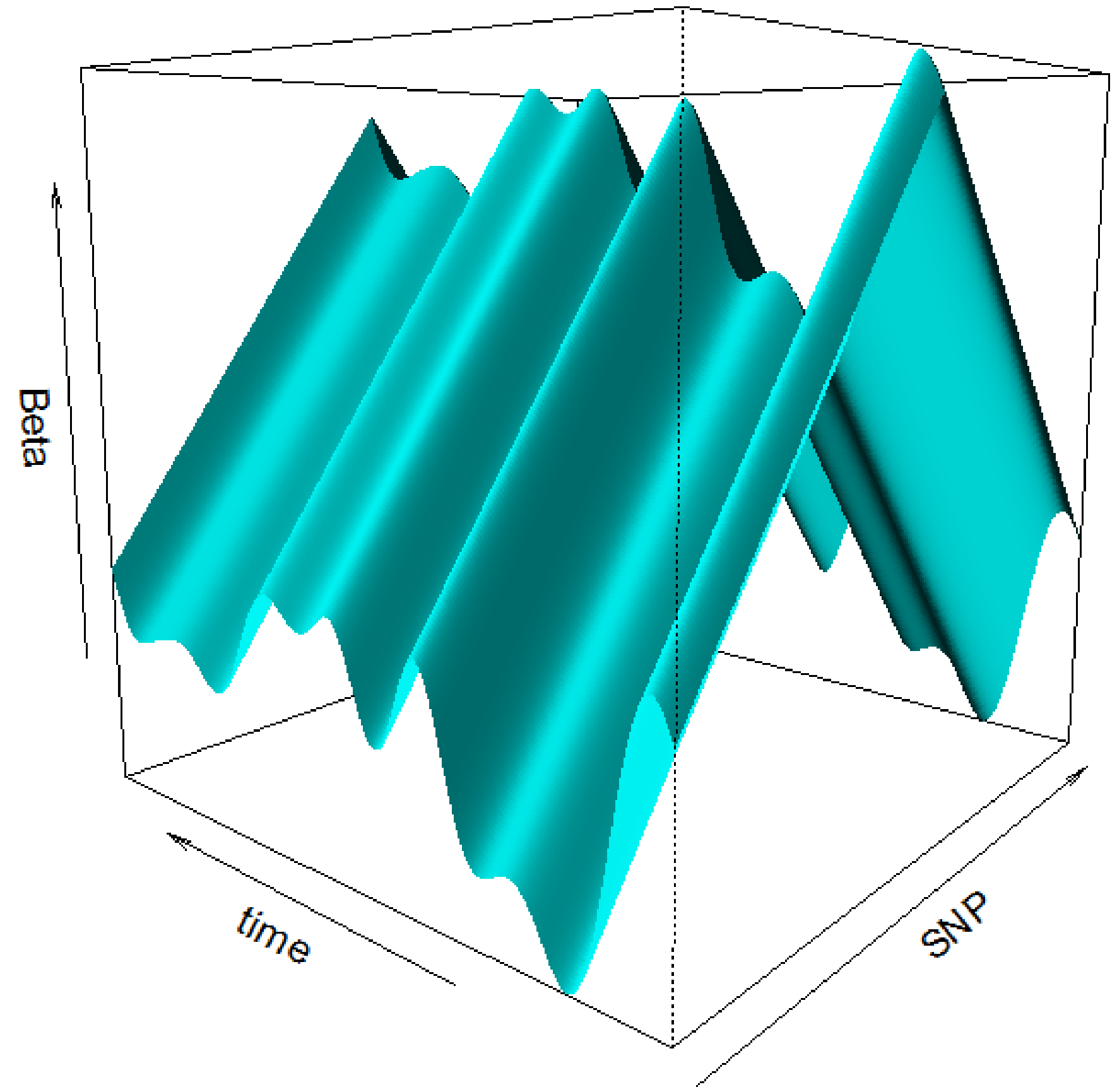

TMEM50B

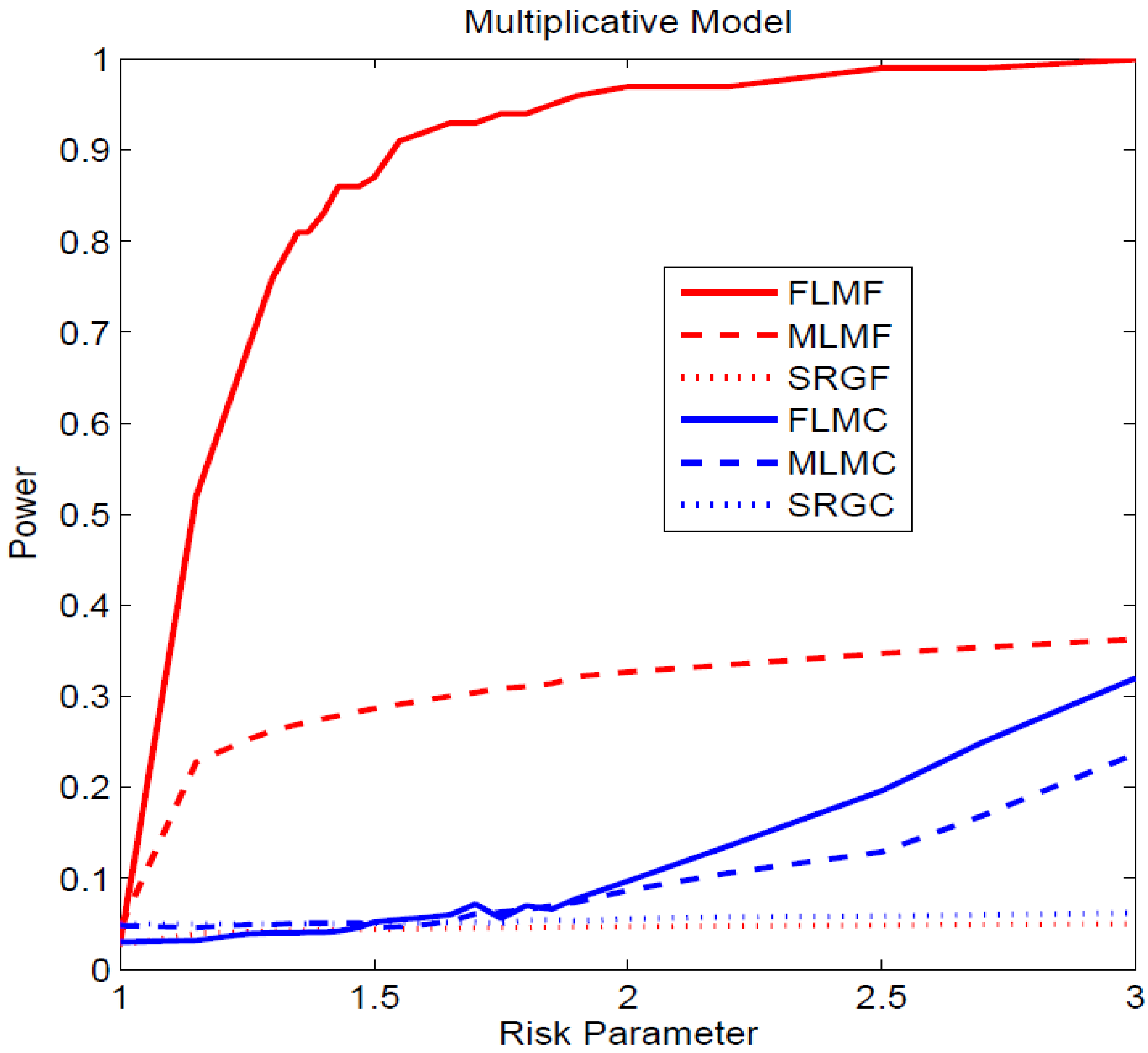

**Figure S1.** The power curves as a function of risk parameter of six models: the functional linear model with both functional response and predictors for function-valued trait (FLMF), the multiple linear model for function-valued trait (MLMF), the simple regression model for function-valued trait (SRGF), the functional linear model with scalar response and functional response for cross section marginal genetic model (FLMC), multiple linear model for cross section marginal genetic model and simple regression for cross section marginal genetic model (SRGC) for testing association of rare variants in the genomic region under multiplicative model at the significance level, assuming a baseline penetrance of 1 and sample sizes of 1,000.

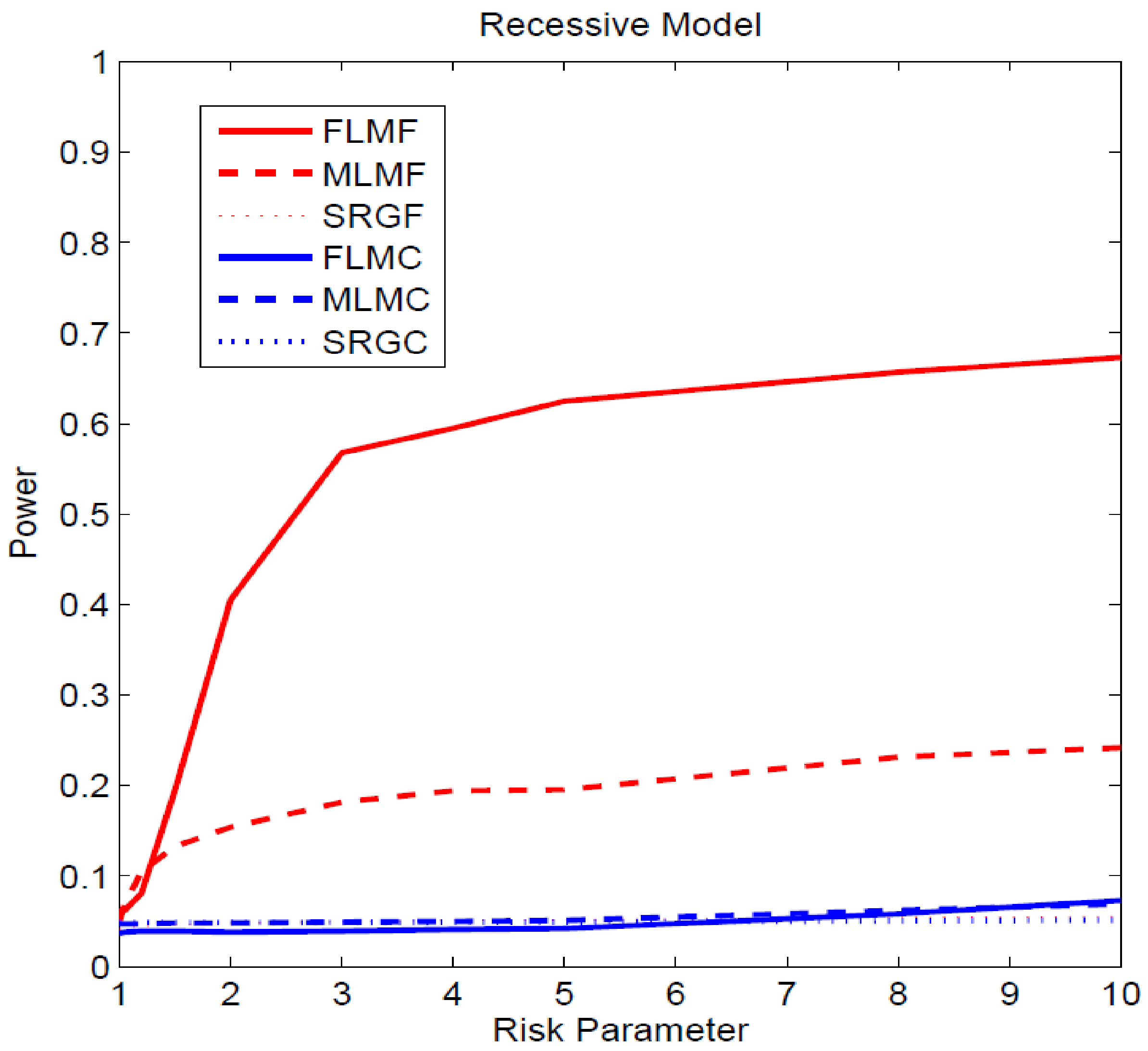

**Figure S2.** The power curves as a function of risk parameter of six models: the functional linear model with both functional response and predictors for function-valued trait (FLMF), the multiple linear model for function-valued trait (MLMF), the simple regression model for function-valued trait (SRGF), the functional linear model with scalar response and functional response for cross section marginal genetic model (FLMC), multiple linear model for cross section marginal genetic model and simple regression for cross section marginal genetic model (SRGC) for testing association of rare variants in the genomic region under recessive model at the significance level, assuming a baseline penetrance of 1 and sample sizes of 2,000.

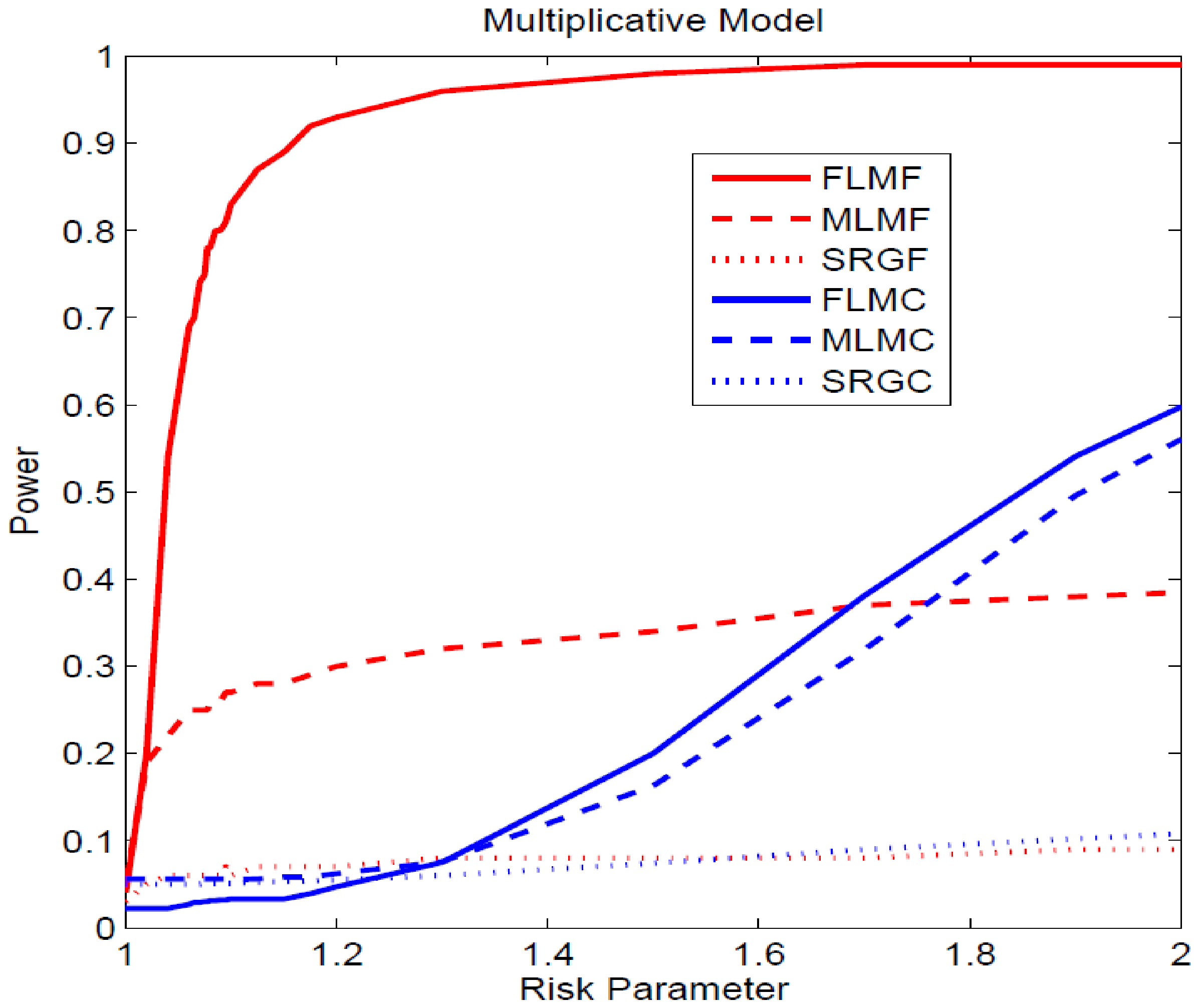

**Figure S3.** The power curves as a function of risk parameter of six models: the functional linear model with both functional response and predictors for function-valued trait (FLMF), the multiple linear model for function-valued trait (MLMF), the simple regression model for function-valued trait (SRGF), the functional linear model with scalar response and functional response for cross section marginal genetic model (FLMC), multiple linear model for cross section marginal genetic model and simple regression for cross section marginal genetic model (SRGC) for testing association of both common and rare variants in the genomic region under multiplicative model at the significance level, assuming a baseline penetrance of 1 and sample sizes of 1,000.

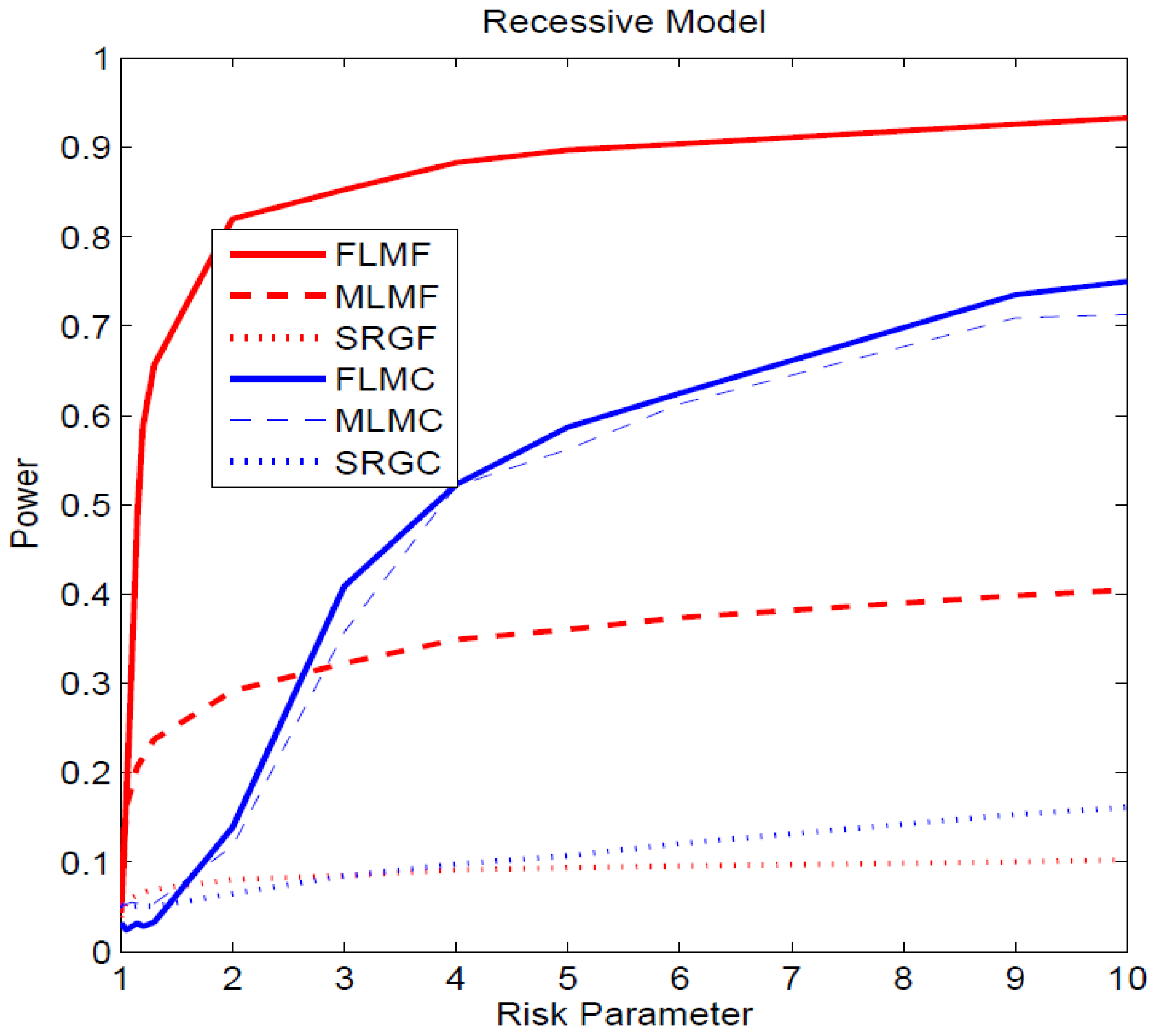

**Figure S4.** The power curves as a function of risk parameter of six models: the functional linear model with both functional response and predictors for function-valued trait (FLMF), the multiple linear model for function-valued trait (MLMF), the simple regression model for function-valued trait (SRGF), the functional linear model with scalar response and functional response for cross section marginal genetic model (FLMC), multiple linear model for cross section marginal genetic model and simple regression for cross section marginal genetic model (SRGC) for testing association of both and rare variants in the genomic region under recessive model at the significance level, assuming a baseline penetrance of 1 and sample sizes of 2,000.

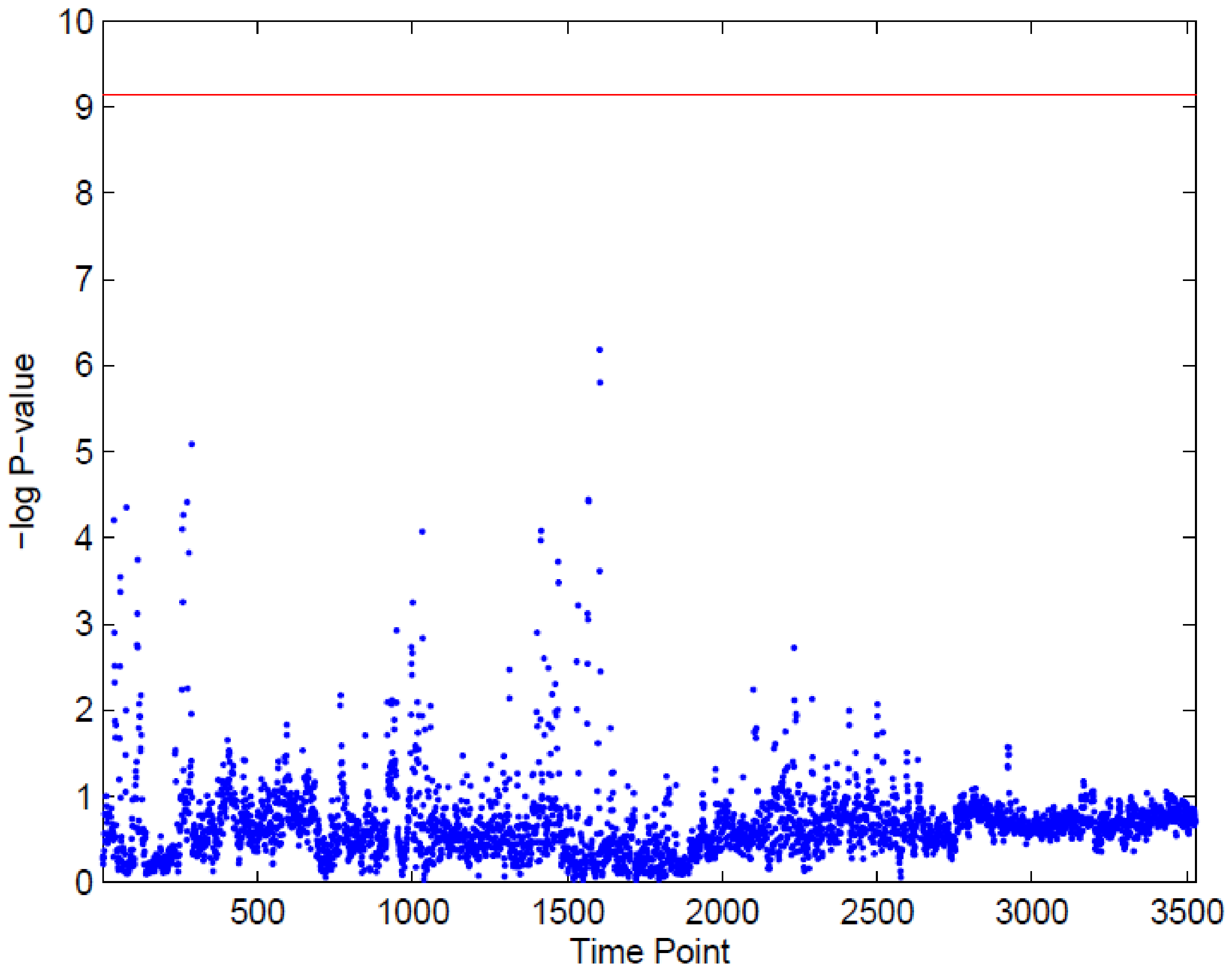

**Figure S5.** –log10 P-value of the MLM for testing the association of gene *ANKLE1* with the oxygen saturation at each time point overnight as a function of time $t$ where the redline indicates the P-value declaring significance.

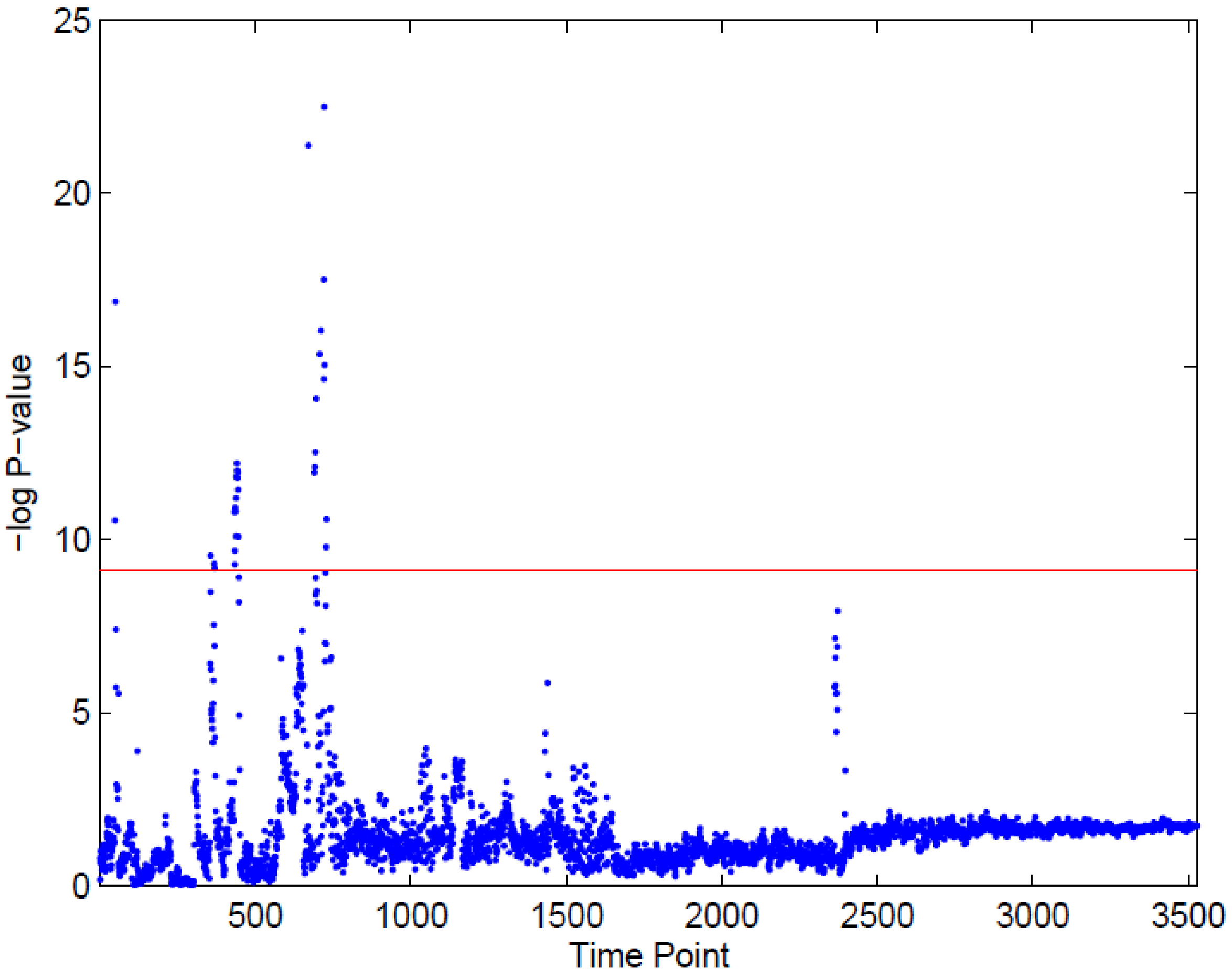

**Figure S6.** –log10 P-value of the MLM for testing the association of gene TMEM50B with the oxygen saturation at each time point overnight as a function of time $t$ where the redline indicates the P-value declaring significance.

Table S1. Distribution of sleep time, age and BMI

| | Male (244) | | | Female (589) | | |
|---|---|---|---|---|---|---|
| | Min | Mean | Max | Min | Mean | Max |
| Average Sleep Time | 4,579 | 22,777 | 33,004 | 2,547 | 22,415 | 33,664 |
| Age | 32.7 | 52.5 | 89.6 | 32.8 | 54.6 | 85.5 |
| BMI | 18.7 | 30.7 | 59 | 20.4 | 32.9 | 60.5 |

Table S2. P-values of 65 significant genes calculated using FLMF, MFMF and SRGF.

| Gene | P-value | | |
|---|---|---|---|
| | FMLF | MLM(min) | SRG(mim) |
| MAN1B1 | 2.53E-21 | 1.10E-05 | 6.57E-03 |
| TMEM57 | 8.90E-18 | 8.50E-02 | 3.97E-02 |
| OR5H15 | 1.28E-17 | 4.21E-03 | 3.29E-02 |
| PABPC4L | 2.66E-15 | 1.09E-01 | 7.95E-02 |
| ANKLE1 | 2.51E-14 | 6.52E-07 | 2.98E-03 |
| TTI2 | 4.64E-14 | 6.82E-02 | 4.57E-03 |
| KRTAP4-7 | 1.67E-13 | 1.12E-02 | 5.45E-03 |
| WDR90 | 1.73E-13 | 1.49E-10 | 4.56E-04 |
| ZER1 | 1.67E-12 | 2.28E-03 | 1.42E-02 |
| DPH2 | 3.05E-12 | 3.09E-03 | 2.12E-02 |
| B9D2 | 3.43E-12 | 1.07E-01 | 2.87E-02 |
| GGT1 | 4.29E-12 | 7.59E-23 | 8.89E-03 |
| SGSH | 4.90E-12 | 3.69E-04 | 4.42E-02 |
| FAM211B | 1.03E-11 | 1.97E-23 | 8.89E-03 |
| FBXO27 | 1.37E-11 | 9.53E-03 | 5.00E-02 |
| COA6 | 1.44E-11 | 1.59E-03 | 2.85E-03 |
| MAK16 | 2.66E-11 | 3.91E-03 | 9.30E-03 |
| CDKN2AIP | 1.69E-10 | 1.62E-01 | 8.54E-02 |
| RRM2 | 3.07E-10 | 6.14E-02 | 1.66E-02 |
| DTX3L | 1.24E-09 | 2.41E-02 | 4.96E-03 |
| C17orf75 | 1.41E-09 | 4.38E-02 | 9.86E-03 |
| TAS2R5 | 1.72E-09 | 2.75E-01 | 4.57E-02 |
| GIPC1 | 1.93E-09 | 2.29E-02 | 4.41E-02 |
| CDC14C | 2.09E-09 | 2.39E-04 | 2.40E-02 |
| MIR4520A | 2.49E-09 | 1.06E-03 | 1.44E-01 |
| MIR4520B | 2.49E-09 | 1.06E-03 | 1.44E-01 |
| PROX2 | 3.63E-09 | 5.24E-03 | 1.15E-02 |
| MAFF | 4.04E-09 | 2.17E-03 | 9.18E-04 |
| UQCRQ | 5.95E-09 | 1.55E-01 | 2.20E-02 |
| LYRM1 | 6.58E-09 | 1.25E-02 | 2.47E-03 |
| ZFPM1 | 7.81E-09 | 2.78E-03 | 9.92E-04 |
| TMEM50B | 9.59E-09 | 3.14E-23 | 1.27E-03 |
| KCNK15 | 1.78E-08 | 1.19E-01 | 7.12E-03 |
| EEF1B2 | 1.99E-08 | 4.60E-03 | 1.08E-03 |
| SNORA41 | 1.99E-08 | 4.60E-03 | 1.08E-03 |
| SNORD51 | 1.99E-08 | 4.60E-03 | 1.08E-03 |
| LDLRAP1 | 2.01E-08 | 3.94E-02 | 2.69E-02 |
| NEK4 | 2.17E-08 | 3.57E-05 | 5.34E-05 |
| COMMD7 | 2.29E-08 | 1.41E-02 | 1.38E-02 |
| EEF1A1 | 3.39E-08 | 3.29E-07 | 1.14E-02 |
| MIR1-1 | 4.51E-08 | 7.91E-02 | 3.37E-02 |
| HMGN4 | 4.63E-08 | 2.05E-03 | 3.47E-03 |
| EVPLL | 5.48E-08 | 3.53E-03 | 1.13E-02 |
| C22orf26 | 5.60E-08 | 1.17E-02 | 1.52E-03 |
| CDC42EP5 | 7.88E-08 | 4.91E-02 | 1.81E-02 |
| MIR29C | 1.30E-07 | 2.86E-02 | 8.72E-03 |
| LHX2 | 1.95E-07 | 6.47E-03 | 2.77E-04 |
| ZNF284 | 2.09E-07 | 1.06E-01 | 7.87E-02 |
| RBAKDN | 2.34E-07 | 2.04E-03 | 3.55E-04 |
| BAIAP2L2 | 2.44E-07 | 1.91E-02 | 2.14E-03 |
| P2RX5 | 2.92E-07 | 6.68E-03 | 1.62E-02 |
| P2RX5-TAX | 2.92E-07 | 6.68E-03 | 1.62E-02 |
| RELB | 3.01E-07 | 1.10E-02 | 1.29E-02 |
| TREML3P | 5.42E-07 | 5.01E-02 | 4.94E-03 |
| TSPAN10 | 5.86E-07 | 5.43E-02 | 5.33E-02 |
| RPS16 | 6.17E-07 | 2.63E-03 | 6.00E-04 |
| GNLY | 8.09E-07 | 1.99E-02 | 1.10E-02 |
| LRRC48 | 8.17E-07 | 1.45E-01 | 1.35E-02 |
| WSB1 | 9.64E-07 | 7.57E-03 | 6.63E-03 |
| GFPT1 | 1.09E-06 | 4.92E-12 | 5.56E-03 |
| MIR3677 | 1.13E-06 | 8.67E-02 | 1.40E-02 |
| MIR940 | 1.13E-06 | 8.67E-02 | 1.40E-02 |
| TOE1 | 1.13E-06 | 6.48E-05 | 6.86E-02 |
| TMEM41A | 2.12E-06 | 5.98E-03 | 3.55E-02 |
| TPM4 | 2.40E-06 | 1.20E-01 | 8.28E-03 |